# EMD-regression for modelling multi-scale relationships, and application to weather-related cardiovascular mortality

Pierre Masselot[1*], Fateh Chebana[1], Diane Bélanger[1,2], André St-Hilaire[1], Belkacem Abdous[3], Pierre Gosselin[1,2,4], Taha B.M.J. Ouarda[1]

[1]*Institut National de la Recherche Scientifique, Centre Eau-Terre-Environnement, Québec, Canada;*

[2] *Centre Hospitalier Universitaire de Québec, Centre de Recherche, Québec, Canada;*

[3]*Université Laval, Département de médecine sociale et préventive, Québec, Canada;*

[4]*Institut national de santé publique du Québec (INSPQ), Québec, Canada.*

*Corresponding Author:*

masselot.pierre@gmail.com

*490, rue de la Couronne*

*G1K 9A9, Québec (QC), CANADA*

July 2017

## Abstract

In a number of environmental studies, relationships between natural processes are often assessed through regression analyses, using time series data. Such data are often multi-scale and non-stationary, leading to a poor accuracy of the resulting regression models and therefore to results with moderate reliability. To deal with this issue, the present paper introduces the EMD-regression methodology consisting in applying the *empirical mode decomposition* (EMD) algorithm on data series and then using the resulting components in regression models. The proposed methodology presents a number of advantages. First, it accounts of the issues of non-stationarity associated to the data series. Second, this approach acts as a scan for the relationship between a response variable and the predictors at different time scales, providing new insights about this relationship. To illustrate the proposed methodology it is applied to study the relationship between weather and cardiovascular mortality in Montreal, Canada. The results shed new knowledge concerning the studied relationship. For instance, they show that the humidity can cause excess mortality at the monthly time scale, which is a scale not visible in classical models. A comparison is also conducted with state of the art methods which are the generalized additive models and distributed lag models, both widely used in weather-related health studies. The comparison shows that EMD-regression achieves better prediction performances and provides more details than classical models concerning the relationship.

# 1. Introduction

In a number of scientific fields (*e.g.* hydrology, environmental health, ecology, etc.), it is of interest to understand the effect of one or several predictor variables on a response variable. The classical class of models for this purpose if regression analysis. However, the variables of interest are often represented by time series processes, which potentially leads to modelling and accuracy issues. The multi-scale nature of some time series processes found in applications such as climatology and public health is of special interest. Indeed, such time series are often non-stationary (*i.e.* the moments vary with time) and some dominant patterns in the time series (*e.g.* annual cycles) create a large amount of multicolinearity in the exposure time series when several covariates are considered. In a regression analysis, if the model does not take these issues into account, it can lead to an increase in the variability of parameter estimates, making the final result less reliable (*e.g.* Ventosa-Santaulària, 2009). This also increases the possibility of making the wrong conclusions concerning whether or not a predictor influences the response (i.e. the so-called "spurious regression" issue, see Granger and Newbold, 1974; Phillips, 1986; Hoover, 2003).

The present paper proposes to address the issue of multi-scale time series data in regression by decomposing the series into intrinsic mode functions (IMF) through the empirical mode decomposition algorithm (EMD, Huang et al., 1998). The obtained IMFs are the basic oscillation modes of time series data, and can be used as variables in a regression analysis. Therefore, the proposed method combines EMD and regression as illustrated in Table 1 and is hereby called "EMD-regression" (EMD-R). The proposed approach differs significantly from other methods commonly used to address the issue of non-stationarity, such as removing the trend and the seasonality (detrending and deseasonalisation), applying a difference operator, or adding a smooth time variable. The main difference lies in the fact that no information is removed from the data. Instead, EMD-R acts as a scan of the relationship over all time scales that are present in the data. This allows isolating the most important time scales for a better understanding of the relationship, and even unveiling signals that may be hidden by the dominant frequencies.

**Table 1: Identification of the proposed EMD-regression method and its link to classic EMD and regression approaches.**

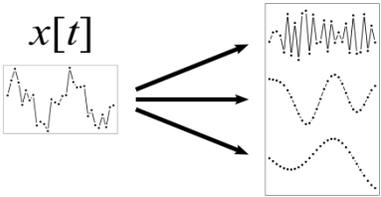

Transforming the data as a prior step to regression analysis has been commonly carried out in the literature, for instance through the use of principal components (e.g. Jolliffe, 1982). More adapted to time series data, a number of spectral decomposition approaches have also been suggested by a number of authors, such as the STL (Seasonal-Trend using Loess) algorithm (e.g. Schwartz, 2000b), Fourier decomposition (e.g. Dominici et al., 2003) or Wavelet transform (e.g. Kucuk and Agiralioglu, 2006; Kişi, 2009). The main advantage of EMD for the decomposition is that it is entirely data-adaptive (Huang et al., 1998). Therefore, the algorithm automatically determines the time scales that are present in the data, avoiding hence the *a priori* choice that is necessary in the STL algorithm (Cleveland et al., 1990) for instance. In addition, no predetermined function is used to perform the decomposition, unlike Fourier and Wavelet based decompositions. This allows EMD to decompose non-stationary and non-linear time series into a small number of components (Huang and Wu, 2008).

In addition to being widely applied directly in several fields such as geosciences (Huang and Wu, 2008) and mechanical engineering (Lei et al., 2013), the EMD algorithm has been successfully combined with other established statistical methods. For instance, Lee and Ouarda (2010); 2011) combined EMD and k-nearest neighbour simulations to predict climatic oscillations. Chen et al. (2012) applied an artificial neural network to forecast the IMFs of a tourism demand series. Lee and Ouarda (2012a) also combined EMD and principal component analysis to separate meaningful signals from noise in climatic applications. EMD has also been used to study the relationship between two variables. For instance, Durocher et al. (2016) used a combination of EMD and cross-wavelet analysis to study the relationship between two time series. For the same purpose, Biswas and Si (2011) and then Hu and Si (2013) used EMD before computing correlation coefficients on the IMFs. A more general method is developed in Chen et al. (2010) to study the correlation between two time series through the use of EMD.

Combining EMD and linear regression has been performed by Yang et al. (2011a) and Yang et al. (2011b). In a recent article, Qin et al. (2016) proposed to use the Lasso approach to select the more relevant IMFs in predicting the response series. The present work goes a step further by proposing a broader scope for the procedure and proposing a number of generalisations of the approach. In particular, the previous studies decomposed only one predictor series, while the present work does not limit itself to only one predictor. In addition, two models are proposed here, one of which decomposes the response series also, allowing its prediction in the frequency space to gain insights at hidden variation scales. A sensitivity score for predictor's IMFs is also described as an interpretation tool for practitioners. Finally, unlike the cited studies, a comparison to state of the art regression methods is provided.

The EMD-regression method basically consists in two steps: i) decomposing the time series into their IMFs through EMD, and ii) using the IMFs as variables in a regression analysis. More specifically, two different designs are introduced: a) only the predictors are decomposed and all their IMFs are used as alternative predictors (EMD-R1) such as in Qin et al. (2016) and b) both the response and predictors are decomposed and each response's IMF is modeled according to the predictors' IMF of the same order (EMD-R2). The new EMD-R2 procedure provides hence more details concerning the relationship between predictors and the response variable than the EMD-R1 procedure.

The present study is motivated by an application in weather-related health, which contains typical examples of multi-scale processes. Such studies often control the seasonality and trend by using a time variable in order to focus on the day-to-day variations in the health issue of interest (Bhaskaran et al., 2013). EMD-regression provides a tool for the assessment of the long term effects of climatic variables through the low frequency IMFs. This represents a major challenge for the planning of future of public health conditions (Xun et al., 2010) and for setting more appropriate public health alerts, especially under climate change conditions. It is hoped that the use of EMD-regression may also unveil hidden features of the weather-health relationship such as the influence of weather factors at non dominant time scales.

The present paper is organised as follows. The background material associated to the EMD-R methodology and the details of the EMD-R approach are introduced in section 2. In section 3, both EMD-R1 and EMD-R2 methods are applied to the weather-related cardiovascular issue in the census metropolitan area (CMA) of Montréal (Canada). Since the motivation context for the present study concerns weather-related health, the EMD-R methods are then compared to commonly used models in this type of study. The results of the application are then discussed in section 4, and the conclusions are presented in section 5.

## 2. EMD-regression (EMD-R)

The EMD-regression methodology aims at explaining the effects of covariates $X_j$ on a response variable $Y$ by: 1) decomposing the time series using EMD and 2) using the IMFs as new variables in a sparse regression model, namely the Lasso (least absolute shrinkage and selection operator, Tibshirani, 1996). The methodology is summarized in Figure 1.

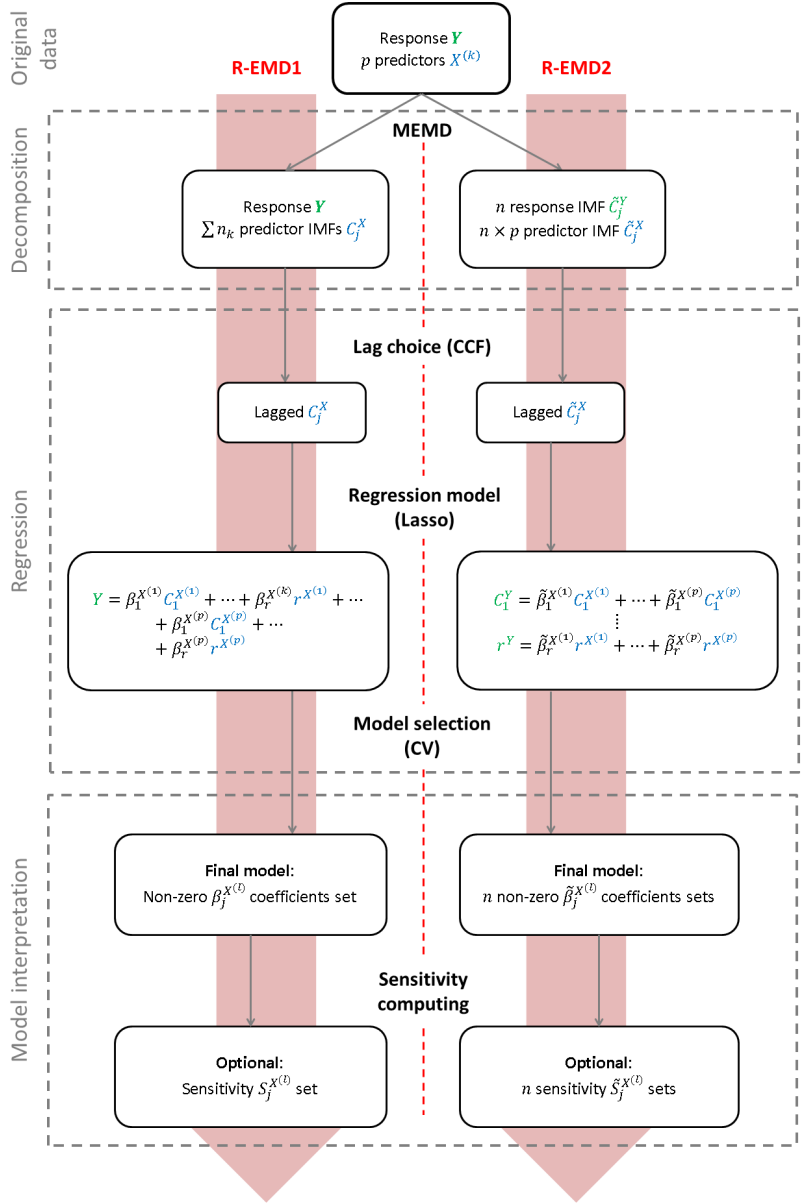

**Figure 1 : Summary of the EMD-regression (EMD-R) methodology.**

## 2.1. Background

The present section introduces the background needed for the two steps of EMD-regression, EMD and the Lasso.

### 2.1.1. Empirical mode decomposition (EMD)

The EMD algorithm (Huang et al., 1998) has been introduced in order to identify a series $X(t)$ oscillation modes $C_k(t)$ without making any *a priori* assumption on the shape of these oscillations. This is achieved by decomposing the series as follows:

$$X(t) = \sum_{k=1}^{K} C_k(t) + r(t) \tag{1}$$

where the $C_k(t)$ $(k=1,...,K)$ are intrinsic mode functions (IMFs), *i.e.* series oscillating around the zero line with symmetric upper and lower envelopes (Huang et al., 1998), and $r(t)$ is a non-oscillating residue. The definition of IMFs allows the $C_k(t)$ s to have meaningful physical interpretation while still representing separate frequency bands.

The IMFs $C_k(t)$ are obtained iteratively, by extracting them from the data beginning with the smallest periodicity to the largest, through a sifting process. The sifting process consists in three steps:

1) obtain the upper and lower envelopes $u(t)$ and $l(t)$ of the data series $X(t)$ by respectively connecting the local maxima and minima through a cubic spline in order to compute their mean $m(t) = \big( u(t) - l(t) \big) / 2$,

2) subtract this mean $m(t)$ to $X(t)$ to obtain the first IMF prototype $h_1(t) = X(t) - m(t)$ and

3) repeat steps 1 and 2 on $h_1(t)$ until obtaining a prototype $h_l(t)$ which corresponds to an IMF, according to some chosen stopping criterion (for different stopping criteria, see Rilling et al., 2003; Huang and Wu, 2008). The prototype $h_l(t)$ is then the first IMF $C_1(t)$ and the whole sifting process is performed again on the residual $r_1(t) = X(t) - C_1(t)$ to obtain $C_2(t)$, then on $r_2(t) = r_1(t) - C_2(t)$ and then again until obtaining a monotonic residual $r_K(t)$. $r_K(t)$ is then considered as the trend of the signal.

When jointly considering $p > 1$ variables, it is useful to obtain the IMFs with the multivariate EMD approach (MEMD, Rehman and Mandic, 2010). It allows to obtain the exact same number of IMFs for each variable, and to obtain mode alignment, *i.e.* two IMFs of the same order from different variables have similar frequency bands (Rehman et al., 2013). The MEMD algorithm works by making a large number of univariate projections of the multivariate series. The envelopes of univariate projections are computed as in the univariate EMD and form together a multidimensional envelope from which a mean can be computed and subtracted from the series (Rilling et al., 2007). Apart from the envelope computation, the MEMD algorithm is identical to the EMD algorithm.

Finally, to avoid mode mixing (i.e. the mixing of very different frequency bands inside one IMF, Huang et al., 1999), the noise-assisted ensemble method is used. It was originally developed by Wu and Huang (2009) for the univariate EMD and later extended by Rehman and Mandic (2011) to the multivariate case (NA-MEMD). The NA-MEMD algorithm adds one or several white noise variables to the multivariate signal in order to make MEMD acting as dyadic filter. Once the MEMD has been performed, the dimensions corresponding to the noise in the resulting multivariate IMFs are discarded. The NA-MEMD is used in the EMD-R procedure but, for simplicity, it is still referred to as EMD throughout the paper.

### 2.1.2. The Lasso

The Lasso (Tibshirani, 1996) is a shrinkage method performing a variable selection while fitting the regression model $Y = \sum_{j=1}^{p} \beta_j X_j + \varepsilon$. The method is specifically designed for high dimensional models, allowing isolating the better predictors among a large set of covariates (e.g. Oloritun et al., 2013). For this purpose, the Lasso has proven more efficient than stepwise algorithms (e.g. Hammami et al., 2012). The Lasso is now very popular in a number of fields, and is integrated as part of many statistical procedures commonly used in various environmental fields (e.g. Bardsley et al., 2015).

The Lasso fits the regression model through the minimization of the penalized least squares criterion

$$\hat{\beta} = \arg\min_{\beta} \left\{ \sum_{i=1}^{n} \left( y_i - \sum_{j=1}^{p} \beta_j X_j \right) + \lambda \sum_{j=1}^{p} |\beta_j| \right\} \tag{1}$$

where $\lambda$ is a tuning parameter controlling the severity of the penalization. The greater $\lambda$ is, the lower the number of covariates $X_j$ remaining in the model at the end is. In practice, $\lambda$ is automatically chosen through cross-validation (for details, see Friedman et al., 2009). Note that the Lasso can also be fit when the response is not Gaussian, similarly to generalized linear models (Park and Hastie, 2007; Friedman et al., 2010).

## 2.2. Proposed EMD-regression approach

As stated in the introduction of this section, the EMD-R methodology contains two designs. In the EMD-R1 design, only the predictors $X_j(t)$ are decomposed and their IMFs $C_{jk}^{(1)}(t)$ are used to explain the original response series $Y(t)$:

$$Y(t) = \sum_{j=1}^{p} \left( \sum_{k=1}^{K} \beta_{jk}^{(1)} C_{jk}^{(1)}(t) + \beta_{jr}^{(1)} r_{jK}^{(1)}(t) \right) + \varepsilon(t) \qquad (2)$$

EMD-R1 is similar as the study of Yang et al. (2011b) and is meant to give an overall view of the relationship.

The EMD-R2 design is meant to provide additional information concerning the relationship at scales with low energy, which are hidden in the original series. In the EMD-R2 design, both the predictors $X_j(t)$ and the response $Y(t)$ are decomposed, leading to the models

$$C_{Yk}^{(2)}(t) = \sum_{j=1}^{p} \beta_{jk}^{(2)} C_{jk}^{(2)}(t) + \varepsilon_k(t) \ \text{ for } \ k = 1, ..., K \qquad (3)$$

for the IMFs and the model

$$r_{YK}^{(2)}(t) = \sum_{j=1}^{p} \beta_{jr}^{(2)} r_{jK}^{(2)}(t) + \varepsilon_r(t) \qquad (4)$$

for the trend. This design is similar to Hu and Si (2013). Since EMD is a complete decomposition (*i.e.* there is not any loss of information in the EMD process), a prediction of $Y(t)$ can be obtained with EMD-R2 by summing up the predictions of the models in (3), *i.e.*

$$\hat{Y}(t) = \sum_{k=1}^{K} \hat{C}_{Yk}^{(2)}(t) + \hat{r}_{YK}^{(2)}(t) \qquad (5)$$

where the $\hat{C}_{Yk}^{(2)}(t)$ and $\hat{r}_{YK}^{(2)}(t)$ are predictions obtained through models (3) and (4). Note that the MEMD allows the number of IMFs $K$ to be constant for all the covariates in both designs (2) and (3), but that the predictors' IMFs $C_{jk}$ can be slightly different in EMD-R1 and EMD-R2, depending on whether the response $Y(t)$ is one of the decomposed variables or not.

In both EMD-R1 and EMD-R2, the Lasso is used to estimate the coefficients $\beta_{jk}^{(m)}$ ($m = 1, 2$). In the case of EMD-R1, it allows to deal with the potentially large number of predictors' IMFs. In the case of EMD-R2, it allows to deal with the potential colinearity between similar frequency predictors' IMFs.

In many applications with time-related data, there may be a delay between a given predictor and the resulting response. Thus, in both EMD-R1 and EMD-R2, the predictors' IMFs $C_{jk}$ may have to be lagged before inclusion in the regression model. The optimal lag should be chosen by maximizing the (absolute) cross-correlation function (CCF) between $C_{jk}$ and the appropriate response (Shumway and Stoffer, 2000). Note that we constrain the lag to be lower than the mean period, computed as the mean interval between two maxima of the corresponding IMF $C_{kj}$ (Wu and Huang, 2004).

It is expected that the variance (or energy) of the IMFs decreases with the frequency (Wu and Huang, 2004). Thus, it may not be convenient to compare directly the $\beta_{jk}$ estimates. A more meaningful quantity to interpret the results is what we hereafter call the sensitivity $S_{jk}$. The sensitivity is the estimated coefficient $\hat{\beta}_{jk}$ standardized by the mean peak-to-peak amplitude $A_{jk}$ of $C_{jk}$, *i.e.*

$$S_{jk} = \hat{\beta}_{jk} \times A_{jk} \qquad (6)$$

The mean peak-to-peak amplitude $A_{jk}$ is used instead of the more traditional standard deviation because of the oscillating nature of $C_{jk}$. Hence, $S_{jk}$ actually indicates the difference that $C_{jk}$ makes in the response when going from a minimum to a maximum value. This quantity is useful to immediately see which oscillating modes mostly influence the response.

## 3. Application to weather-related cardiovascular mortality

The literature abounds with studies documenting the potentially harmful impacts of climate change. Among these impacts, it is expected to observe an increase in weather-related mortality. Cardiovascular diseases (CVD) are among the diseases that are most affected by climate change since they are impacted by extreme weather (e.g. Braga et al., 2002; Bustinza et al., 2013). CVD are already the main cause of mortality in Canada and could represent an increasing burden on the Canadian public health system in future years (Wielgosz et al., 2009). Therefore, it is important to properly understand the impact of weather on CVD, in order to organize more efficiently adaptation strategies and reduce the eventual harmful effects of climate change. To help achieving this objective, EMD-R is hereby applied to the study of weather-related cardiovascular mortality in the city of Montreal (Canada). The first sub-section introduces the data and the second presents the results.

### 3.1. Data

The data used are from the Montreal's CMA (geographical location shown in Figure 2). This region represents the densest population basin in the province of Quebec, allowing enough CVD death cases in a relatively small area. This allows the model to be relevant since weather can be considered homogeneous within this small and flat area.

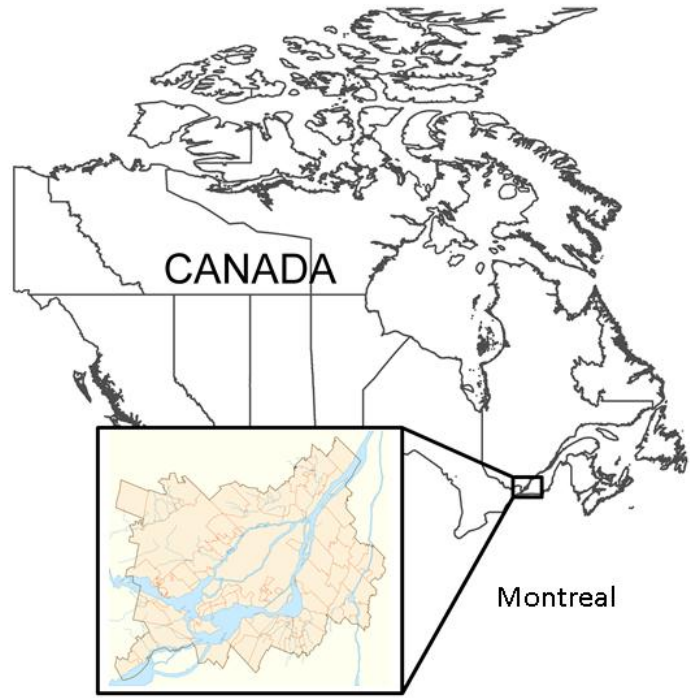

**Figure 2: Map and location of the study region, i.e. the greater Montreal in the province of Quebec, Canada.**

In the present application, the response series ($Y$) is the mortality ($M$) series, *i.e.* the daily number of CVD deaths from 1981 to 2011 (a total of $n = 11322$ days). CVD includes ischaemic heart diseases (I20-I25 in the tenth version of the international classification of diseases, ICD-10), heart failure (I50 in the ICD-10), cerebrovascular diseases and Transient cerebral ischaemic attacks (G45, H34.0, H34.1, I60, I61, I63 and I64 in the ICD-10).

The predictor series ($X_j$s) are daily weather variables in the same days than the CVD death data. To illustrate the fact that the EMD-R methodology can be applied when several predictors are considered, temperature ($T$) and relative humidity ($H$) are chosen in the present application. Temperature represents the most studied predictor in environmental epidemiology (Gasparrini et al., 2015) and humidity is also sometimes considered as a predictor (Schwartz et al., 2004) or as a confounding variable (Yang et al., 2015; Phung et al., 2016). Weather data series are measured at several stations located over the Greater Montreal area. The final series used for the whole area is the simple arithmetic mean of all the station series. Note that, in a previous study, weighted kriging was considered instead of spatial mean, but it was concluded that it did not improve the models' performance (Giroux et al., 2013).

## 3.2. Results

This section presents the results obtained by applying the EMD-R model described in section 2 on the data introduced in section 3.1. In the following, an IMF will be denoted by $C_{Xf}^{(m)}$ for a given variable $X$ (*i.e. M*, *T* or *H* respectively for mortality, temperature and humidity), *f* is its mean period in days and *m* represents the model ($m = 1$ means that we are in the EMD-R1 context and $m = 2$ represents the EMD-R2). The mean periodicity *f*, computed as the mean difference between two successive peaks in the IMF (Wu and Huang, 2004) is more useful than the order *j* for the interpretation of results (although the two are related since the order *j* increases with the periodicity *f*). For instance, $C_{T365}^{(1)}$ is the temperature IMF representing the annual cycle (*i.e.* a periodicity of 365 days) used in the EMD-R1 model. This change in notation also impacts the quantities associated to the IMFs such as the amplitudes, regression coefficients and sensitivities now respectively denoted $A_{Xf}^{(m)}$, $\beta_{Xf}^{(m)}$ and $S_{Xf}^{(m)}$.

For the present application, the parameters of the NA-MEMD model are set based on Rehman et al. (2013). The stopping criterion for the sifting process is the one of Rilling et al. (2003) and two white noise variables with a variance equal to 10% of the variance of the data are added to perform NA-MEMD. The number of projections for computing the multidimensional envelopes is set to 128.

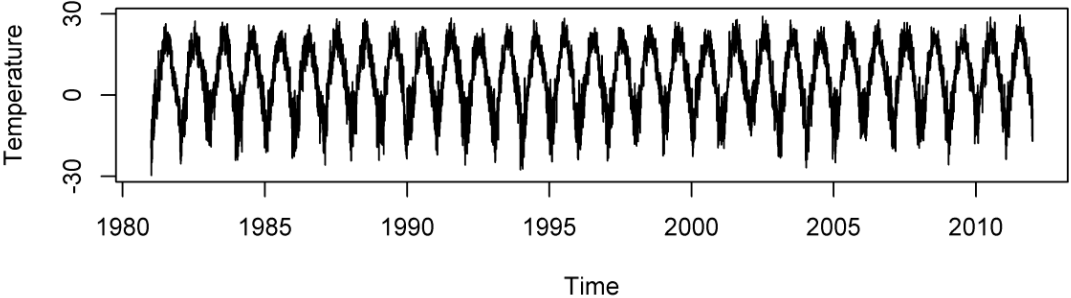

a) Original series

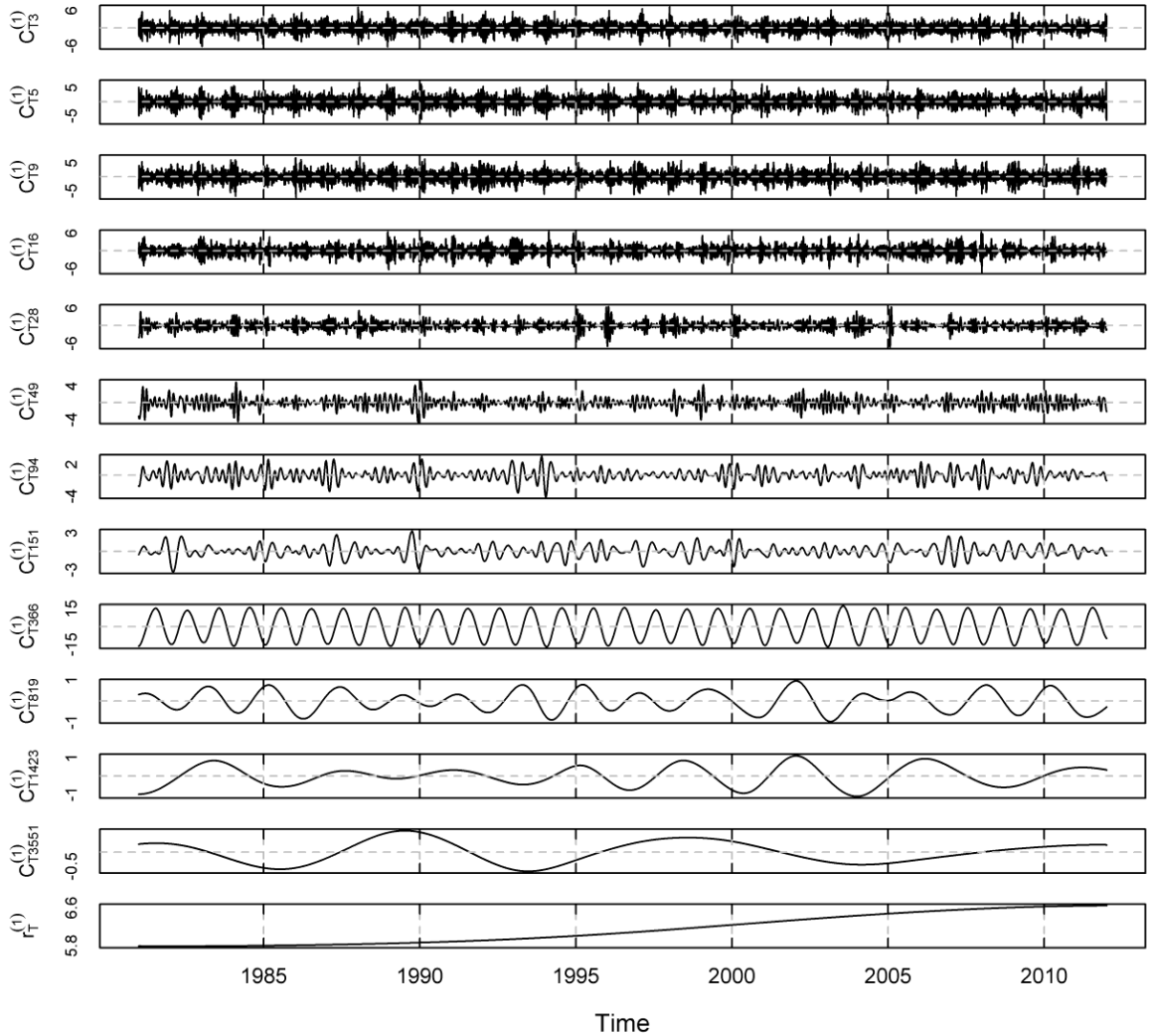

b) Decomposed series

**Figure 3: Temperature series (panel a) and IMFs $C_{Tf}^{(1)}$ obtained by MEMD for the EMD-R1 model (panel b).**

### 3.2.1. Interpretation of the results

To provide an example of EMD decomposition, Figure 3 shows the temperature series of the Greater Montreal (Figure 3a) and its IMFs $C_{Tf}^{(1)}$ obtained by MEMD to be used as predictors in the EMD-R1 model (Figure 3b). It can be seen that the frequency of each IMF is indeed regular but that the amplitude may vary inside one IMF. Figure 3 confirms that EMD results in components that can be interpreted, since the extremely regular annual cycle as well as the increasing trend induced by climate change, are well represented. The IMFs shown in Figure 3, along with the humidity IMFs $C_{Hf}^{(1)}$ are then the predictors of the EMD-R1 model.

The sensitivities $S_{Xf}^{(m)}$ estimated by the EMD-R1 and EMD-R2 models (2) and (3) are shown in Figure 4. Since theoretical confidence intervals cannot be computed with the Lasso approach (Lockhart et al., 2014), the 95% confidence intervals (CI) are computed through 500 moving-block bootstrap replications (Chatterjee and Lahiri, 2011). Note that, the "one standard error rule" in cross-validation for choosing the parameter $\lambda$ in (1) has been used in the case of EMD-R1 (e.g. Krstajic et al., 2014). This rule takes the uncertainty of the cross-validation into account, and allows obtaining the sparsest model possible since the goal of EMD-R1 is to depict an overall image of the relationship.

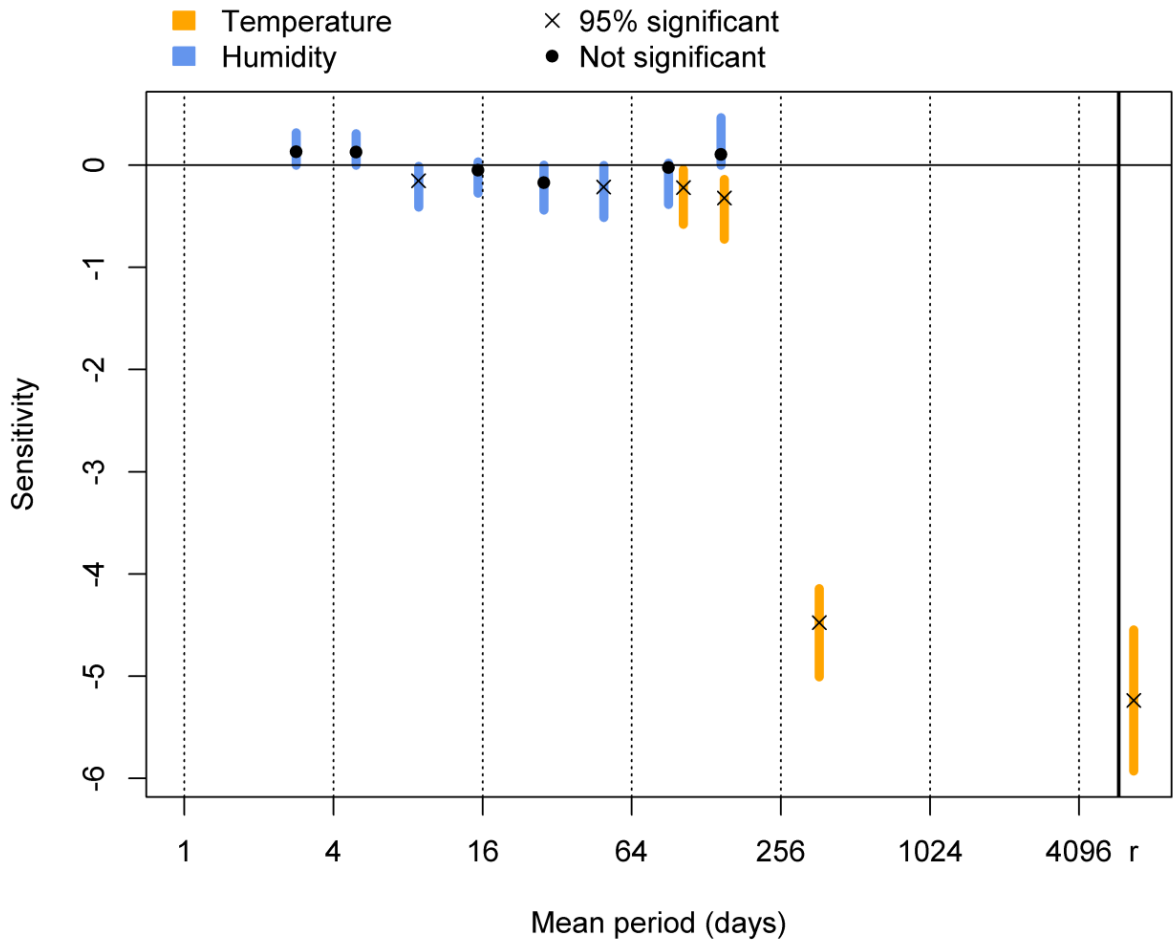

a)  EMD-R1

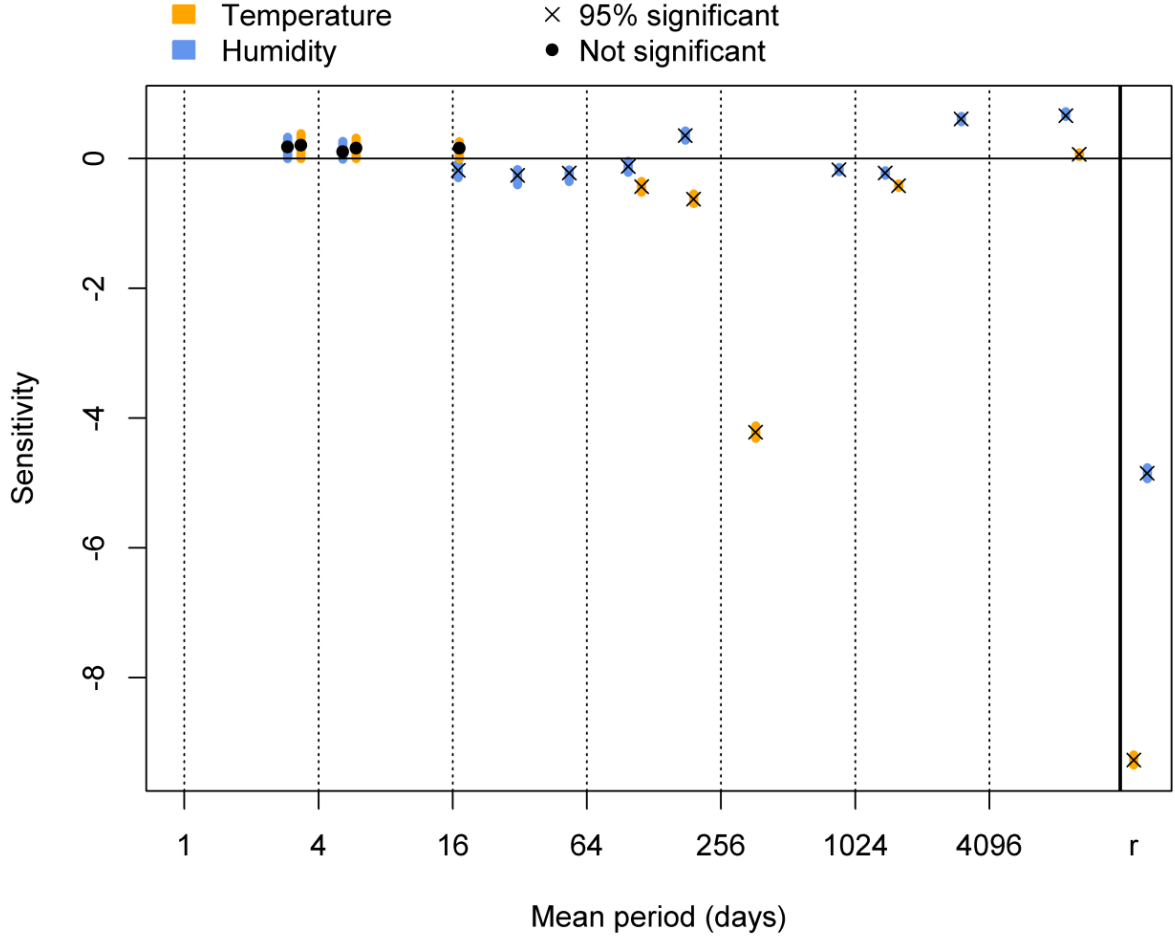

b) EMD-R2

**Figure 4: Non-null sensitivities $S_{Xf}^{(.)}$ obtained by EMD-R1 (panel a) and EMD-R2 (panel b) according to the mean period of the associated IMF $C_{Xf}^{(.)}$.**

**The label "r" in the right area of the x-axis represents the trend of the series (the residue component in equation (1)). The black points are the Lasso estimates on the whole dataset while the segments indicate 95% confidence intervals obtained using 500 block-bootstrap replications.**
**"x"s represent the Lasso estimations for which the confidence interval does not reach the zero line, which can be seen as the "significant" coefficients. Only the non-null coefficients (*i.e.* kept by the Lasso) are shown. The x axis is in binary logarithmic scale because of the dyadic nature of EMD.**

A comparison of the results of EMD-R1 (Figure 4a) and EMD-R2 (Figure 4b) shows that they are similar, only with more details for EMD-R2. Focusing on Figure 4a indicates that there is a clear separation between humidity (in blue), which affects the mortality at short time scales, and temperature (orange) which affects the mortality at large time scales. At the shortest time scales, humidity presents positive sensitivities; but we note that in this case the CIs reach the zero line, meaning a lack of confidence about this effect. However, at slightly larger time scales (*i.e.* $C_{H9}^{(1)}$ and $C_{H49}^{(1)}$), there are sensitivities between -0.15 and -0.20 with CIs not containing the zero value. Cumulating them indicates that there is an extra death every 3 days during the dryer periods compared to the humid ones. Figure 5 shows the mean amplitude of $C_{H9}^{(1)}$ and $C_{H49}^{(1)}$ in one year (the top two panels) and indicates that their amplitude during the months of March to May is twice the amplitude of the remaining months of the year. Therefore, the dry effect on mortality is more important during spring season.

According to Figure 4, temperatures have an effect on mortality only at periodicities larger than 3 months. There is especially the obvious annual effect with $S_{T365}^{(1)} = -4.47$ meaning that there is an increase of more than 4 deaths a day when $C_{T365}^{(1)}$ is at its minimum (*i.e.* winter) in comparison to when it is at its maximum (*i.e.* summer). Aside from the annual effect, there are also slight negative effects of $C_{T94}^{(1)}$ and $C_{T151}^{(1)}$. Figure 5 (two bottom panels) indicates that these IMFs have their highest amplitudes during winter which means that they aggravate the already strong winter effect. Finally, the strongest effect reported by Figure 4 is the effect of the increasing temperature trend (*i.e.* the global warming), which is associated to the decreasing trend in cardiovascular mortality. Therefore, it would seem that the global warming tends to mitigate the winter effect.

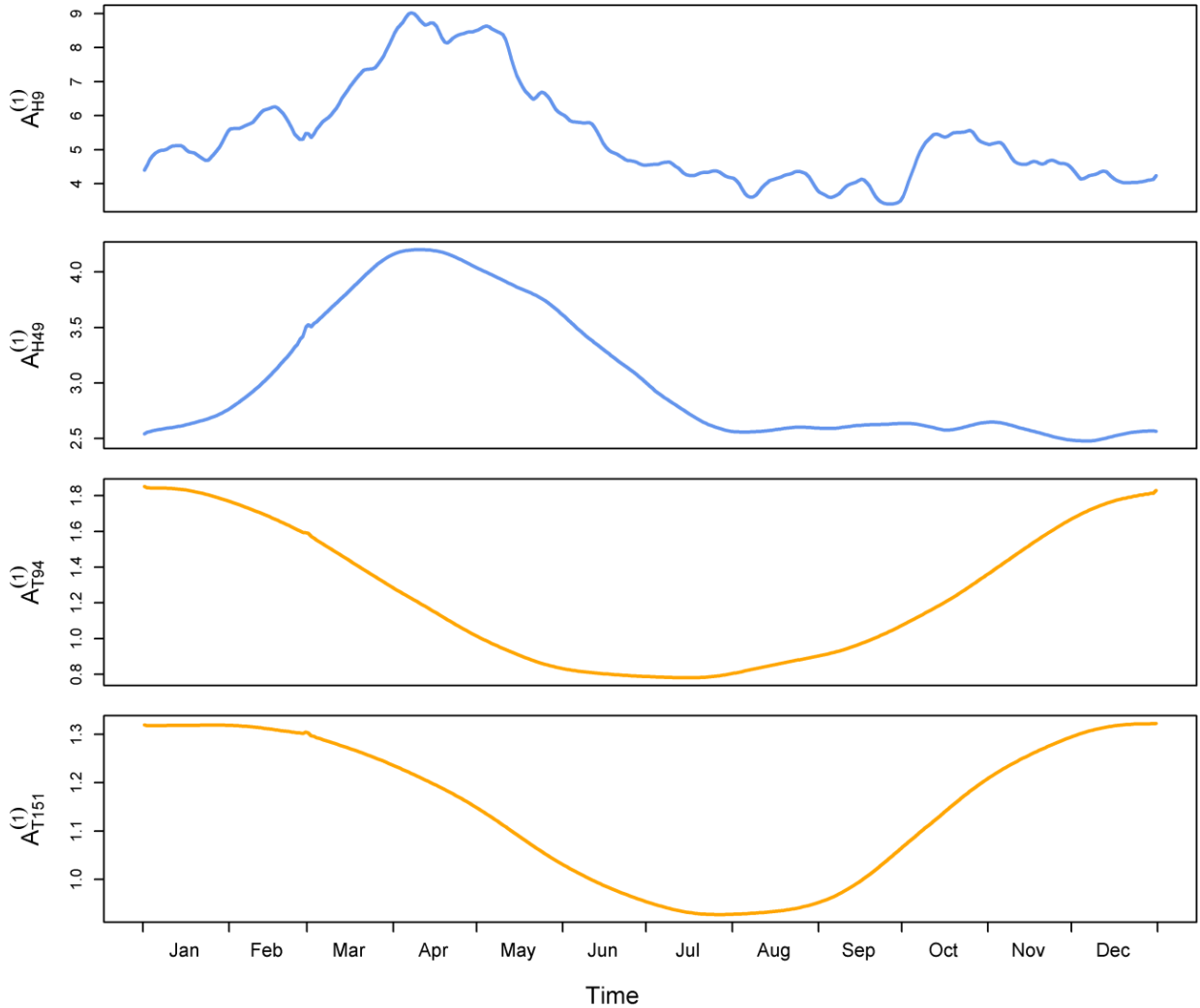

In addition to the results of EMD-R1, EMD-R2 gives more insights on the relationship. Figure 4b shows positive sensitivities for the IMFs corresponding to the shortest temperature periodicities ( $C_{T3}^{(2)}, C_{T5}^{(2)}$ and $C_{T17}^{(2)}$ ), but with CI reaching the zero line meaning that we are not too confident about this effect. The biggest difference between EMD-R1 and EMD-R2 results, is that the latter indicates effects at very large time scales. There are especially high sensitivities for humidity at the 8-year and 25-year scales ( $S_{H3059}^{(2)} = 0.35$ and $S_{H9017}^{(2)} = 0.66$ ). Finally, Figure 4b also indicates that the humidity trend $r_H^{(2)}$ combines with the temperature trend $r_T^{(2)}$ to explain the mortality trend.

This application has been purposely kept simple to focus on the methodology itself. However, in more complete epidemiologic studies, some variables such as age and gender must be controlled. Models for different age classes and for both genders have been performed in the present case study. Nevertheless, no real differences have been observed from the general results presented above in this section. In addition, models dealing with separating winter and summer have also been performed since the effect of weather can change according to the season. The obtained results do not provide much added value to the present paper. Hence, they were not presented here to limit the complexity of the application. Further models and results on this topic are presented in the extensive application performed in the technical report of Masselot et al. (2015).

### 3.2.2. Performance assessment and comparison

The previous section has shown a qualitative difference between EMD-R1 and EMD-R2, *i.e.* the latter retains more weather IMFs as predictors and hence, more information than the EMD-R1. However, EMD-R1 is easier to interpret because it is more parsimonious. The present section aims at quantitatively comparing the two EMD-R designs. In addition, EMD-R is compared to the most widely used regression models in weather-related health studies: generalized additive models (GAM, Hastie and Tibshirani, 1986) and distributed lag nonlinear models (DLNM, Gasparrini et al., 2010).

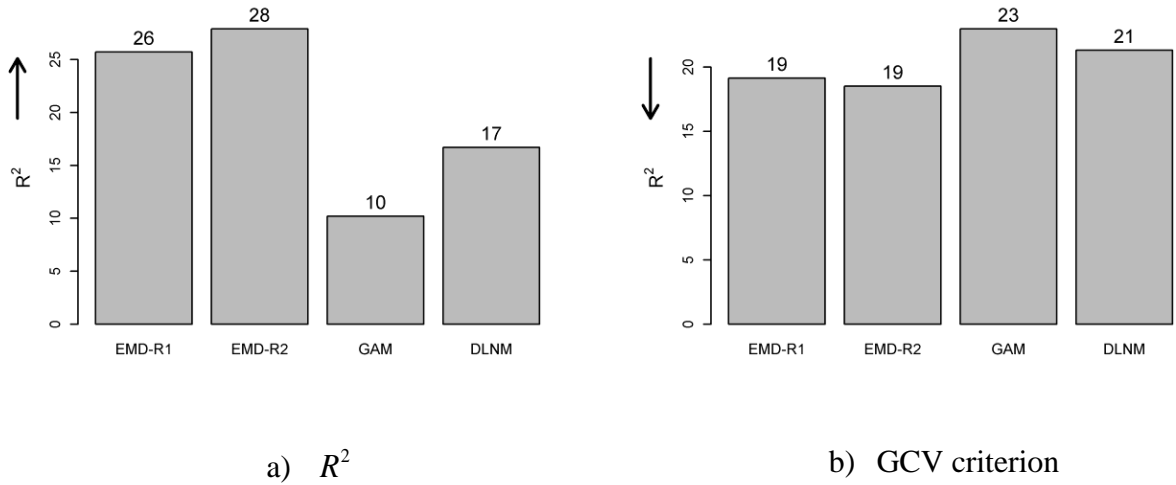

a)  $R^2$                  b)  GCV criterion

**Figure 6: Comparison of the performance criteria $R^2$ and GCV of the two EMD-R models as well as GAM and DLNM applied on the same data.**

GAM are commonly used in environmental epidemiology for having brought to attention the nonlinear J, U or V-shape relationships between mortality and temperatures (e.g. Braga et al., 2001; Bayentin et al., 2010). DLNMs are also nonlinear models, adding the effect of delayed predictors (e.g. Li et al., 2013; Wu et al., 2013; Vanos et al., 2014). The strength of both models lies in their capacity to model the different effects of hot and cold temperatures. However, these models focus on day-to-day variations to estimate the relationship, while the application of EMD-R shows that the relationship also exists at different scales than the daily one.

The numerical comparison between the models is performed through the in-sample $R^2$ and generalized cross-validation (GCV, Craven and Wahba, 1978) criteria. The $R^2$ measures the ability of a regression model to fit the observed data. However, since a high $R^2$ can indicate overfitting, the GCV is also computed to measure the prediction ability of regression models. The GCV is an estimation of the prediction error of the model and often gives estimations similar to leave-one-out cross-validation.

The values of the $R^2$ and GCV criteria for each model are shown in Figure 6. One can see that the EMD-R1 and EMD-R2 models display the best $R^2$ values with 26% and 28% respectively, while GAM and DLNM have $R^2$ values of 10% and 16% respectively (Figure 6a). The GAM and DLNM $R^2$ values are consistent with the values usually reported in the literature. The higher score for EMD-R2 shows that this model is more accurate than EMD-R1, since its details allow explaining a larger proportion of the response's variance. Note that these $R^2$ scores are particularly high knowing that the weather is actually one of many factors (and not the main one) affecting the CVD mortality. Other factors include, for instance, physical exercise, obesity, dietary habits and smoking (Institut national de santé publique du Québec, 2006).

The GCV scores (Figure 6b) lead to the conclusion that EMD-R models have lower prediction error than GAM and DLNM. However these differences are not large differences since GAM and DLNM have scores of 23 and 21 versus scores of 19 for both EMD-R1 and EMD-R2.

## 4. Discussion

The results of the weather-related cardiovascular mortality presented in section 3, already show one advantage of the EMD-R: its ability to display some hidden aspects of the relationship. In this case, the effect of humidity found during spring season and at very large time scales (*i.e.* periodicities of several years) is quite new in the field of environmental epidemiology. Indeed, no significant association between relative humidity and mortality has been found when studying as a variable of interest with classical models (e.g. Schwartz et al., 2004).

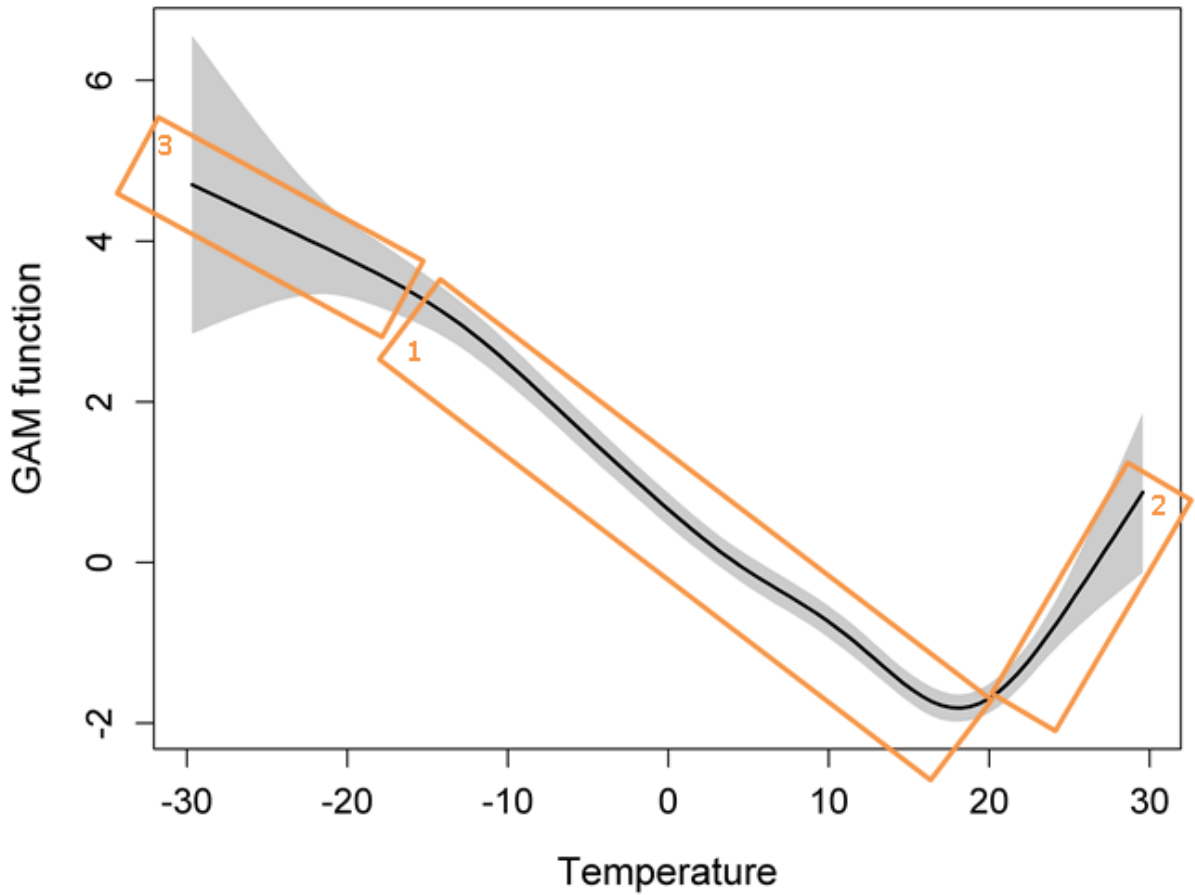

Environmental epidemiology studies often concentrate on temperatures (e.g. Patz et al., 2014). In section 3.2, the effect of temperature is mainly an effect of winter cold weather. The readers from the field of environmental epidemiology may be surprised about the non-significance of short periodicity temperature IMFs, since the effect of heat waves is the most documented one in the literature (e.g. Chebana et al., 2012; Bustinza et al., 2013). However, heat waves are extreme events, not necessarily well represented by single IMFs, and need particular statistical methods to be studied. Hence, depending on the objective, it may be necessary to complete this analysis with extreme values models. The present finding that the main effect is the constant cold is actually consistent with the global study of Gasparrini et al. (2015).

In addition to the association with humidity, the results of the application unveil the potential influence of climate on cardiovascular mortality. Indeed, the long term association found with EMD-R2 could be a first clue of teleconnections. Therefore, EMD-R can be a tool for studying the long-term influence of climate change, which is an important challenge in environmental health (Xun et al., 2010). This is also a known challenge in the field of air pollution (Koop and Tole, 2006). Such studies could use the recently developed tools for forecasting future oscillations of the IMFs (Kurbatskii et al., 2011; Lee and Ouarda, 2012b; Ghouse et al., 2015). It could be of interest to use or adapt these tools to forecast the predictors' IMF in order to anticipate the evolution of CVD mortality related to climate change. Note that all the tools commonly used in the field of environmental epidemiology, such as the computation of relative risks and the use of a time variable to control for unmeasured confounders, can be used in the context of EMD-R.

The previous section shows a large amount of results because the relationship between weather and cardiovascular mortality is a complex one, being nonlinear with different lags depending on the season and the location (*e.g.* Gasparrini et al., 2015). Other cases could be simpler such as the relationship between health and atmospheric pollutants, often considered linear with relatively short lags (see, e.g., the review of Brook et al., 2010). However, the performance comparison of section 3.2.2 shows that EMD-R offers an improvement over reference methods in the field of environmental epidemiology, both for explanation and prediction. At first thought, this might seem odd since GAM and DLNM are both nonlinear while the EMD-R is only linear, and that the relationship between temperatures and mortality is well known for being nonlinear. This is illustrated by Figure 7 which shows the J-shaped GAM function obtained for temperatures on the cardiovascular mortality for the data described in section 3.1. However, note that this function is piecewise linear and that all the pieces can be found in Figure 4b. Indeed, the coefficient $\hat{\beta}_{T365}^{(2)} = -0.14$ corresponds to the main piece in box 1 which has a slope of approximately -0.15. Moreover, the positive coefficients $\hat{\beta}_{T3}^{(2)}$, $\hat{\beta}_{T5}^{(2)}$ and $\hat{\beta}_{T17}^{(2)}$ (although not significant) correspond to the increasing piece of box 2 and the negative coefficients $\hat{\beta}_{T94}^{(2)}$ and $\hat{\beta}_{T151}^{(2)}$ correspond to the slightly decreasing piece of box 3. The usefulness of nonlinear models is that they summarize several effects in one curve but EMD-R is able to provide details about the different parts of a nonlinear relationship.

Despite its usefulness, EMD-R is not intended to be an alternative to the most used regression methods but instead as a complement to their results. Indeed, EMD-R results are difficult to interpret "at a glance" like, *e.g.* GAM functions. They should be treated as explorative results giving clues for further research and their results should be compared with those of other methods. An issue is that an uncertainty is associated to the estimation of IMFs especially those corresponding to the low frequency (Huang et al., 2003). For example, the sensitivities found for low frequency IMFs cannot lead to final conclusions but must be investigated further. An interesting option could be to compare the sensitivities with the results of the IMF significance test (Wu and Huang, 2004) to determine if there is indeed a signal. Another alternative would be to use an error-in-variable regression to include the IMFs' error directly in the model (*e.g.* Fuller, 2009).

Another methodological limitation lies in the choice of lags. Indeed, the present methodology chooses only one lag per IMF, but in many cases like in environmental epidemiology, the lags are distributed, meaning that the effect of an exposure is not only on a single day (Schwartz, 2000a). Therefore, a perspective would be to develop a statistical method able to estimate distributed lags as well as perform a variable selection, for instance a mix between the group Lasso (Yuan and Lin, 2006) and the lag weighted Lasso (Park and Sakaori, 2013).

## 5. Conclusion

The present paper introduces a general methodology for EMD-regression when dealing with time series data (and more generally all data with autocorrelation) often found in environmental sciences. The purpose of the EMD-R approach is to understand a relationship between variables from a different point a view, *i.e.* from a time scale point of view. This point of view acknowledges the complexity of many real-world time series which contain a significant amount of information in their variations. With EMD-R, it is possible to assess the influence of all of these time scales on the response variable of interest, while the classical regression will only depict the relationship at the dominant time scale. In addition, the use of the Lasso to perform the regression analysis in both EMD-R1 and EMD-R2 allows considering several predictors in the analysis. Aside from the core of the EMD-R methodology (*i.e.* EMD and Lasso), the present paper also proposes tools that help in the interpretation, as illustrated in the application results. These tools are the sensitivity analysis (6) with an associated plot (illustrated in Figure 4) and the amplitude plot (illustrated in Figure 5).

Applying the EMD-R methodology to environmental epidemiology data leads to unprecedented results. Indeed, the present results strongly suggest an effect of humidity on mortality, which is not established in the public health literature. The innovative results illustrate the relevancy of applying the methodology in a variety of fields.

The present paper intended to introduce the EMD-R in its basic form but for a general context to show its applicability and benefits in fields where regression is used and data are time dependent. EMD-R can be used in a number of fields such as hydrology, ecology, geophysics, etc. Therefore, as a perspective, EMD-R can be used in the study of weather-related cardiovascular mortality in a more complex framework that controls some other variables, such as atmospheric pollutants. From a statistical perspective, it could also be of interest to study the impact of IMF estimation error on the EMD-R final results.

## Acknowledgements

The authors are thankful to the Fonds Vert du Québec for funding this study and to the Institut national de santé publique du Québec for data access. The authors also thank Jean-Xavier Giroux (INRS-ETE) for his help on the database establishing as well as Yohann Chiu (INRS-ETE) for all his relevant comments during the project. The authors are grateful to Scott Sheridan, the associate editor of *Science of the total environment* as well as three anonymous reviewers for their judicious comments and help improving the readability of the paper. The analyses have been performed with the open source software R (R Core Team, 2015) for which a package implementing the tools proposed in the present paper is available.